\documentclass[conference]{IEEEtran}
\IEEEoverridecommandlockouts
\usepackage{cite}
\usepackage{amsmath,amssymb,amsfonts,amsthm}
\usepackage{algorithm}
\usepackage[noend]{algpseudocode}
\usepackage{graphicx}
\usepackage{tabularx}
\usepackage{textcomp}
\usepackage{breqn}
\usepackage{xcolor}
\def\BibTeX{{\rm B\kern-.05em{\sc i\kern-.025em b}\kern-.08em
    T\kern-.1667em\lower.7ex\hbox{E}\kern-.125emX}}
\usepackage{pgfplots}

\usepackage{tikz}
\usetikzlibrary{arrows, arrows.meta, patterns,positioning,shapes.geometric}
\usepackage{dsfont}
\usepgfplotslibrary{fillbetween}

\newtheorem{theorem}{Theorem}
\newtheorem{lemma}{Lemma}
\newtheorem{definition}{Definition}

\newtheorem{cor}{Corollary}

\newcommand{\ceiling}[1] {\left\lceil #1 \right\rceil}
\newcommand{\floor}[1]{\left\lfloor{#1}\right\rfloor}
\newcommand{\setof}[1]{\left\{{#1}\right\}}

\newcommand{\len}{\ell en}
\newcommand{\Tbb}{\mathbb{T}}
\newcommand{\eqdef}{:=}

\begin{document}

%\title{Response-Time Analysis for Probabilistic Conditional Parallel DAG Tasks}
\title{Response-Time Analysis and Optimization for Probabilistic Conditional Parallel DAG Tasks}

\author{
\IEEEauthorblockN{Niklas Ueter}
\IEEEauthorblockA{\textit{TU Dortmund University}} 
Dortmund, Germany \\
niklas.ueter@tu-dortmund.de
\and
\IEEEauthorblockN{Mario G\"unzel}
\IEEEauthorblockA{\textit{TU Dortmund University}} 
Dortmund, Germany \\
mario.guenzel@tu-dortmund.de
\and
\IEEEauthorblockN{Jian-Jia Chen}
\IEEEauthorblockA{\textit{TU Dortmund University}} 
 Dortmund, Germany \\
jian-jia.chen@tu-dortmund.de
}

\maketitle
\thispagestyle{plain}
\begin{abstract}
 Real-time systems increasingly use multicore processors 
 in order to satisfy thermal, power, and computational requirements.
 To exploit the architectural parallelism offered by the multicore processors,
 parallel task models, scheduling algorithms and response-time analyses with 
 respect to real-time constraints have to be provided. 
 In this paper, we propose a reservation-based scheduling algorithm for 
 sporadic constrained-deadline parallel conditional DAG tasks with probabilistic 
 execution behaviour for applications that can tolerate bounded number of deadline 
 misses and bounded tardiness. 
 We devise design rules and analyses to guarantee 
 bounded tardiness for a specified bounded probability for $k$-consecutive deadline misses 
 without enforcing late jobs to be immediately aborted. 
\end{abstract}


\section{Introduction}
\label{sec:introduction}
A real-time system is a system where the missing 
of a deadline may lead to a catastrophe and thus warrants to formally verify the temporal 
behaviour of the system to ensure safety.
In the last decade real-time systems have shifted from uniprocessor to multiprocessor 
systems in order to deal with the computational, thermal and energy constraints 
of modern complex applications.
To that end, a lot of research has been conducted
with regards to the challenge of how to make use of the parallelism provided by 
multiprocessors for task sets with inter- and intra-task parallelism whilst 
satisfying deadline constraints. Inter-task parallelism refers to the potential 
concurrent execution of distinct tasks that execute sequentially, whereas intra-task parallelism 
refers to tasks that allow for parallel execution. 
Fork/join models~\cite{Lakshmanan:2010:SPR:1935940.1936239}, synchronous parallel task models, 
real-time scheduling algorithms and response-time analyses thereof have been published, e.g.,~\cite{SaifullahRTSS2011}, and DAG (directed-acyclic graph) based task models~\cite{Fonseca:2017:RTNS, DBLP:conf/sies/FonsecaNNP16,  DBLP:conf/ipps/Baruah15, DBLP:conf/ecrts/BonifaciMSW13,MelaniECRTS2015}.
These models enable tasks with higher execution demands and inherent 
parallelism such as computer vision, radar tracking or video applications 
to be scheduled with tighter deadlines.

Besides the different approaches and justifications to represent
intra-task parallelism using the above models, parallel applications in 
the domain of autonomous driving and image processing are subject to 
multiple conditional branches and control flow instructions as stated by Melani et. al~\cite{MelaniECRTS2015}.
Moreover, the execution times of the subjobs of parallel algorithms
in these domains are highly varying due to varying sensor inputs, e.g., 
images for object detection in autonomous vehicles.
Beyond that, it was shown that the multicore architecture complicates the 
worst-case timing analysis. This is due to interference effects from contention 
on shared resources, e.g., caches, memory etc. 
The authors in~\cite{DBLP:conf/wcet/FernandezAQRVC14} argue that the \emph{arbitration delay} 
and \emph{state perturbation} caused by resource sharing must be captured in the 
worst-case bounds. 
All these uncertainties eventually lead to pessimistic response-time analyses 
in real-time systems and thus lead to resource underutilization.
These architectural impacts on the worst-case execution time analysis 
have been thoroughly researched by 
e.g., cache partitioning~\cite{Altmeyer2014} or bandwidth sharing mechanisms for 
memory accesses~\cite{Yun2013}.

Another approach to this problem is to \emph{accept} the uncertain 
execution behaviour of the parallel tasks and to focus on the 
probabilistic response-time characteristics. 
For many applications, e.g., closed-loop feedback controllers, hard real-time 
system engineering (with a safe but very pessimistic upper bound) is not required due to the inherent controller robustness 
towards timing non-idealities like jitter and deadline misses.
In fact, if only a limited number of deadlines of a control application are missed, 
the required quality of control can still be satisfied. 

Recently, many research efforts have been focused on formalizing 
and analyzing relaxations of deadline constraints~\cite{DBLP:conf/ecrts/PazzagliaMMC19}, e.g., 
weakly hard systems
where $m$ out of $k$ task instances must meet the deadlines.
Moreover, Maggio et al.~\cite{maggio_et_al:LIPIcs:2020:12384} investigate 
the closed-loop control system stability under consecutive deadline-miss constraints, 
which further motivates the need for scheduling algorithms that can guarantee probabilistic 
bounds on consecutive deadline misses to the application.

In order to formally describe and verify quantitive guarantees of
deadline misses, some quantifications are of importance for soft
real-time systems: Probability of a deadline miss, probability for $k$
consecutive deadline misses, maximum tardiness of a job.  Despite the
guarantees are soft, the precise quantification of such deadline
misses are hard and challenging even for the ordinary sequential
real-time task models that are scheduled upon a uniprocessor system. A
summary of the literature in this research direction is provided in
Section~\ref{sec:related-work}. They can only be derived under
strict model assumptions, e.g., that a job is aborted whenever a job
exceeds its deadline in the state-of-the-art analyses.  The reason for this complexity is partly due
to inter task interference, i.e., the preemption and interference
patterns of the task system due to higher-priority jobs, which results
in a large number of system states that must be considered in a
response-time analysis.

We aim to analyze, optimize and verify the schedulability of probabilistic conditional parallel DAG tasks 
on identical multi-processors with respect to quantities such as deadline-miss probabilities, 
consecutive deadline-miss probabilities and tardiness constraints.
When considering the scheduling and analysis of probabilistic parallel DAG tasks, 
not only inter-task, but also intra-task interference, and multiprocessor scheduling anomaly 
effects (the early completion of jobs may lead to longer response-times) 
must be considered, which complicate the analyses for the above 
mentioned quantities.

\noindent{\bf Contributions}:
We propose scheduling algorithms based on reservations, 
i.e., service provisioning, for the probabilistic analysis of parallel DAG tasks to 
avoid inter-task interference induced complexities and anomaly effects and are thus 
firstly able to solve the stated objective.
More precisely, we make the following contributions:
\begin{itemize}
\item We propose a probabilistic version and formal description of the widely 
 used conditional parallel DAG task model in Section~\ref{sec:task-and-problem-model}.

\item We contribute scheduling algorithms and response-time analyses
  for probabilistic conditional parallel DAG tasks based on resource
  reservation.  The reservations can be scheduled along side real-time
  workloads using any existing scheduling paradigm. In addition, we
  provide design rules to devise reservations that guarantee
  probabilistic characteristics such as bounded tardiness, stability,
  and probabilistic upper-bounds for $k$-consecutive deadline misses.
  Our approach is anomaly-free because any early completions due to
  scheduling or dynamic DAG structures are handled by the adoption of
  resource reservation and the abstraction of the workload model.
\end{itemize}

To the best of our knowledge, this is the first paper that addresses
the analysis and optimization for probabilistic conditional parallel
DAG task sets with quantitive guarantees.

\section{Related Work}
\label{sec:related-work}

The scheduling of parallel real-time tasks with 
worst-case parameters, e.g., worst-case execution times, 
upon multiprocessor systems has been extensively studied for 
different parallel task models. An early classification of 
parallel tasks with real-time constraints into \emph{rigid}, \emph{moldable} or \emph{malleable} 
has been described by Goosens et al.~\cite{DBLP:journals/corr/abs-1006-2617}.
Early work concerning parallel task models focuses on synchronous parallel task models, 
e.g.,~\cite{DBLP:conf/rtns/MaiaBNP14, SaifullahRTSS2011, DBLP:conf/ecrts/ChwaLPES13}. Synchronous models are an extension of the fork-join model~\cite{Conway63} 
in the sense that it allows different numbers of subtasks in each (synchronized) segment 
and that this number could be greater than the number of processors.
Many of the proposed scheduling algorithms and analyses are based on decomposition, 
i.e., the decomposition of the parallel task into a set of sequential tasks 
and the scheduling thereof.

Recently, the directed-acyclic graph (DAG) task model has been proposed 
and been subject to scheduling algorithm design and analysis.
The DAG task is a more general parallel structure where each task 
is described by a set of subtasks and their precedence constraints 
that are represented by a directed-acyclic graph. 
This parallel model has been shown to correspond to models in parallel computing APIs 
such as OpenMP by Melani et al.~\cite{DBLP:conf/cases/SerranoMVMBQ15} 
or Sun et al.~\cite{DBLP:conf/rtss/SunGWHY17}.
This model has been studied in the case of global 
scheduling in e.g.,~\cite{DBLP:conf/ecrts/BonifaciMSW13,nasri_et_al:LIPIcs:2019:10758} or partitioned scheduling algorithms~\cite{DBLP:conf/sies/FonsecaNNP16, 8603232}.
There has also been research regarding approaches of synchronous and 
general DAG tasks that are not decomposition based, e.g., federated 
scheduling as proposed by Li et al.~\cite{Li:ECRTS14} that avoids inter-task 
interference for parallel tasks.
In federated scheduling, the set of DAG tasks are partitioned into 
tasks that can be executed sequentially on a single processor 
whilst meeting it's deadline requirements and tasks that need 
to execute in-parallel in order to meet it's deadline. 
The latter tasks are then assigned to execute on a 
set of processors exclusively.

Motivated by the conditional execution behaviour of modern parallel applications, 
e.g., autonomous driving or computer vision, the conditional DAG task model 
has been proposed. A plethora of research concerning the real-time schedulability 
of this model has been conducted by e.g., \cite{MelaniECRTS2015, DBLP:conf/emsoft/Baruah15, ChenPeng2019}.
Most recently, the computational complexity of the scheduling of conditional DAG with 
real-time constraints has been investigated by Marchetti et al.~\cite{DBLP:conf/ipps/Marchetti-Spaccamela20}.
However, due to the worst-case parameters and the worst-case conditional structure 
that has to be considered during real-time verification of the scheduling algorithms, 
resource over-provisioning is inevitable. 

For soft real-time applications that can tolerate a bounded number of deadline-misses, 
probabilistic task models and response-time analyses for these kind of parallel tasks 
are of interest. Moreover, the worst-case parameter inference is increasingly 
complex and pessimistic for parallel architectures further bolstering the importance 
of probabilistic models and analyses.
For sequential stochastic tasks a plethora of prior work concerning 
probabilistic analyses exists, 
e.g.,~\cite{DBLP:conf/etfa/SantinelliYMC11, DBLP:conf/rtns/HobbsTA19}.
Recent work focused on the improvements of efficiency in convolution-based 
probabilistic deadline-miss analysis approaches.
In Br\"uggen et al.~\cite{DBLP:conf/ecrts/BruggenPCCM18}, the authors propose 
efficient convolutions over multinomial distributions by exploiting 
several state space reduction techniques and approximations 
using Hoeffding's and Bernstein's inequality and unifying equivalence classes.
Chen et al.~\cite{khchenRTCSA18} propose 
the efficient calculation of consecutive deadline-misses using 
Chebyshev's inequality and moment-generating functions and 
optimizations thereof.
There has also been efforts to use reservation servers to schedule 
probabilistic sequential tasks.
For example, Palopoli et al.~\cite{DBLP:journals/tpds/PalopoliFAF16} 
have shown how to calculate the probability of a deadline miss for 
periodic real-time tasks scheduled using the constant bandwidth server (CBS).
The authors have reduced the computation to the computation of a steady state 
probability of an infinite state discrete time markov chain with periodic 
structure.
In the context of parallel DAG tasks Ueter et al.~proposed a reservation 
scheme to schedule sporadic arbitrary-deadline DAG tasks~\cite{Ueter2018} 
with real-time constraints. 
Other approaches to tackle the probabilistic 
analysis of real-time tasks is real-time queuing theory by Lehoczky et al.~\cite{Lehoczky96}, 
which is an extension of classical queuing theory to systems with deadlines.
An initial work that analyzed the probabilistic response-times 
of parallel DAG tasks was proposed by Li~\cite{LiECRTS14Stochastic}.
Li extended prior work on federated scheduling~\cite{Li:ECRTS14} 
by facilitating queuing theory to devise federated scheduling 
parameters such that each task's tardiness is bounded and soft 
real-time requirements are met.
A more recent work on the probabilistic response-time analysis of parallel DAG tasks 
is by Ben-Amor et al.~\cite{Amor2019, Amor2020}.
The authors have studied the probabilistic response-time 
analysis of parallel DAG tasks upon multiprocessor systems using partitioned 
fixed-priority scheduling at the subtask level. 
In their model each subtask is described by a probabilistic worst-case execution 
time and static precedence constraints between them. 
Based on the above, the authors derive probabilities for subtask 
response-times using convolution-based approaches and compose an 
overall response-time.

\section{Task and Problem Model}
\label{sec:task-and-problem-model}

\begin{figure}[tb]
\centering
\begin{tikzpicture}[]
\def\ux{1.3cm}\def\uy{0.35cm}
\tikzset{>={Latex[width=1.5mm,length=1mm]}}
        \tikzset{
           job/.style={minimum height=0.1*\uy, circle, draw, inner sep=2pt},
		  decision/.style={fill=black, diamond, outer sep=4pt},
        }
    
\node[job, anchor=south west, label=below:{$v_1$}, draw] at (0, 0)  (C1) {$3$};
\node[job, anchor=south west, label=below:{$v_2$}, draw] at (2*\ux, -2*\uy)  (C2) {$1$};
\node[job, anchor=south west, label=below left:{$v_3$}, draw] at (2*\ux, 0.5*\uy)  (C3) {$2$};
\node[job, anchor=south west, label=below:{$v_4$}, draw] at (2*\ux, 3*\uy)  (C4) {$1$};
\node[job, anchor=south west, label=below:{$v_5$}, draw] at (4.5*\ux, -1*\uy)  (C5) {$2$};
\node[job, anchor=south west, label=below:{$v_6$}, draw] at (4.5*\ux, 2*\uy)  (C6) {$5$};
\node[job, anchor=south west, label=below:{$v_7$}, draw] at (5.5*\ux, 0.5*\uy)  (C7) {$3$};
	
\node[diamond, anchor=south west, fill=black, draw, scale=0.4] at (1*\ux, 2*\uy)  (D1) {};
\node[diamond, anchor=south west, fill=black, draw, scale=0.4] at (3.5*\ux, 1*\uy)  (D2) {};

\path[->]	(D2)		edge	[]	node [below] {\small $0.4$}	(C5);
\path[->]	(D2)		edge	[]	node [above] {\small $0.6$}	(C6);

\path[->]	(C2)		edge	[]	node [above] {}	(C3);
\path[->]	(C5)		edge	[]	node [above] {}	(C7);
\path[->]	(C6)		edge	[]	node [above] {}	(C7);

\path[->]	(C1)		edge	[]	node [above] {}	(C2);
\path[->]	(C1)		edge	[]	node [above] {}	(D1);
	
\path[->]	(C2)		edge	[]	node [above] {}	(D2);

\path[->]	(C4)		edge	[]	node [below] {}	(D2);
\path[->]	(C3)		edge	[]	node [above] {}	(D2);
        
\path[->]	(D1)		edge	[]	node [above] {\small $0.7$}	(C4);
\path[->]	(D1)		edge	[]	node [below] {\small $0.3$}	(C3);

    \end{tikzpicture}
\caption{An exemplary probabilistic conditional DAG task in 
 which each conditional node (diamond) denotes that only one 
 of it's adjacent subjobs is released (with the annotated probability) during runtime.
 In this specific example four different DAG structures can be instanced during runtime.}
\label{fig:example-conditional-dag-task}
\end{figure}
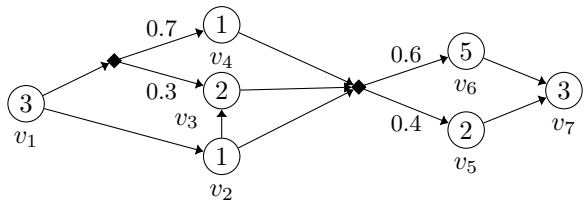

We consider a given set $\mathbb{T}$ of probabilistic sporadic constrained-deadline 
conditional parallel directed-acyclic graph (DAG) tasks in a multiprocessor system that is comprised 
of $M$ identical (homogeneous) processors. 
Each task releases an infinite sequence of task instances, namely 
jobs. Each conditional parallel DAG task $\tau_i \in \mathbb{T}$ is defined by a 
conditional DAG structure $G_i$ (to be defined later), a relative deadline $D_i$ and a 
minimal inter-arrival time $T_i$, which denotes the minimal distance 
between two job releases. In this paper we only consider constrained-deadline 
tasks, i.e., $D_i \leq T_i$ for every task $\tau_i$. 
An exemplary probabilistic conditional DAG 
is illustrated in Figure~\ref{fig:example-conditional-dag-task}.
A probabilistic conditional directed-acyclic graph is composed of 
nodes $V$ and edges $E$ that denote precedence and control flow constraints. 
Each node is either a \emph{subjob} node with an associated 
execution time or a \emph{condition} node that denotes probabilistic conditional 
branching to subjobs.
In the illustrated example, two decision nodes with two possible branching options each 
are given. The given structure yields four different enumerable DAG realizations 
whose probability of realization is given by the probability of traversing a specific path 
of condition nodes. 
A conditional DAG is composed of finitely many DAGs, 
each of which consist of a tuple $(V, E)$, where $V$ denotes the finite set 
of subjobs and the relation $E \subseteq V \times V$ denotes the precedence constraints 
of these subjobs such that there are no directed circles in the underlying graph.
For each of these DAGs the \emph{volume} and \emph{length} parameters are calculated as 
follows.
We use $pre(v_{i}) \eqdef \{v_{j} \in V~|~(v_{j}, v_{i}) \in E\}\text{\;\:and\;\:}v_{j} \prec v_{i}\text{~if~}v_{j} \in pre(v_{i})$
Conversely, we use $succ(v_{i}) \eqdef \{v_{j} \in V~|~(v_{i}, v_{j}) \in E\}\text{\;\:and\;\:}v_{j} \succ v_{i}\text{~if~}v_{j} \in succ(v_{i})$.

\begin{definition}[Path]
\label{def:path}
 A path $\pi$ in a directed-acyclic graph $G$ is any sequence of subjobs 
 $v_{i_1} \prec v_{i_2} \prec \ldots \prec v_{i_k}$ for $v_{i_j}\in V$ 
 such that $pre(v_{i_1}) = \emptyset$ and $succ(v_{i_k}) = \emptyset$.
\hfill \IEEEQED
\end{definition}

\begin{definition}[Length]
\label{def:dag-length} 
 Let a path $\pi$ be a sequence of subjobs 
 such that each subjob in the sequence is an immediate successor of the 
 previous subjob in terms of precedence constraints. 
 Then the length of a path is given by 
 \[ \ell en(\pi) \eqdef \sum_{v_{i} \in \pi} \ell en(v_{i}) \] where the length of a subjob 
 denotes its execution time. Subsequently, the length of DAG $G$ is given by 
 \[ \ell en(G) \eqdef \max \{ \ell en(\pi)~|~\pi~is~a~path~in~G \}. \]
\hfill \IEEEQED
\end{definition}

\begin{definition}[Volume]
\label{def:dag-volume}
 The volume of DAG $G$ is given by the graph's cumulative execution time, 
 i.e., \[ vol(G) \eqdef \sum_{v_{i} \in V} \ell en(v_{i}). \]
\hfill \IEEEQED
\end{definition}

\subsection{Probabilistic Parametric Description}
\label{sec:parametric-description}

\begin{table}[tb]
\centering
\begin{tabular}{c c c}
{\bf\textsc{probability}} & {\bf \textsc{length}} & {\bf \textsc{volume}} \\
 		0.42 & 12 & 13 \\ 
 		0.18 & 13 & 14 \\ 
 		0.28 & 9 & 10\\ 
 		0.12 & 11 & 11\\ 
\end{tabular}
\vspace{0.8em}
\caption{Tabular representation of the probabilities of the parameters volume and length 
 for the probabilistic conditional DAG task illustrated in Figure~\ref{fig:example-conditional-dag-task}.}
\label{table:example-probabilities}
\end{table}

Each probabilistic conditional DAG task is described by the tuple $\tau_i = (G_i, D_i, T_i)$ 
where $G_i$ denotes a probabilistic conditional DAG structure, $D_i$ denotes the relative deadline and 
$T_i$ denotes the minimal inter-arrival time between two job releases. 
For each task $\tau_i \in \mathbb{T}$ a cumulative distribution function (CDF) 
is inferred from the conditional DAG structure,  where $F_i(u, v)$ describes the probabilistic behaviour of the 
\emph{volume} and \emph{length} of a DAG instance.
That is each task $\tau_i$ releases an infinite number of jobs 
$\tau_{i, \ell},~\ell=0,1,2, \dots$ and each job is associated with a DAG instance $G_{i, \ell}$ such that the 
parameters \emph{volume} and \emph{length} of $G_{i, \ell}$ are a realizations according to 
the probabilistic characterization of the distribution function. 

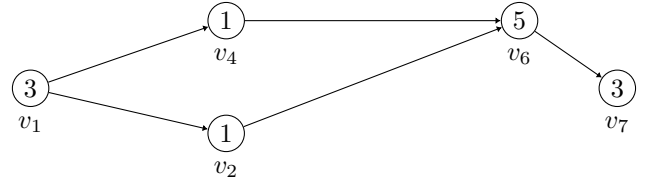
\begin{figure}[tb]
\centering
\begin{tikzpicture}[]
\def\ux{1.3cm}\def\uy{0.3cm}
\tikzset{>={Latex[width=0.9mm,length=0.7mm]}}
        \tikzset{
           job/.style={minimum height=0.1*\uy, circle, draw, inner sep=2pt},
		  decision/.style={fill=black, diamond, outer sep=4pt},
        }
    
\node[job, anchor=south west, label=below:{$v_1$}, draw] at (0, 0)  (C1) {$3$};
\node[job, anchor=south west, label=below:{$v_2$}, draw] at (2*\ux, -2*\uy)  (C2) {$1$};

%\node[job, anchor=south west, draw] at (2*\ux, 0.5*\uy)  (C3) {$2$};
\node[job, anchor=south west, label=below:{$v_4$}, draw] at (2*\ux, 3*\uy)  (C4) {$1$};

%\node[job, anchor=south west, draw] at (5*\ux, -2*\uy)  (C5) {$2$};
\node[job, anchor=south west, label=below:{$v_6$}, draw] at (5*\ux, 3*\uy)  (C6) {$5$};

\node[job, anchor=south west, label=below:{$v_7$}, draw] at (6*\ux, 0)  (C7) {$3$};

\path[->]	(C6)		edge	[]	node [above] {}	(C7);
\path[->]	(C1)		edge	[]	node [above] {}	(C4);
\path[->]	(C1)		edge	[]	node [above] {}	(C2);
\path[->]	(C2)		edge	[]	node [above] {}	(C6);
\path[->]	(C4)		edge	[]	node [below] {}	(C6);

    \end{tikzpicture}
\caption{DAG instance of the exemplary conditional DAG task shown in Figure~\ref{fig:example-conditional-dag-task} where 
the conditional branches with probability $0.7$ and $0.6$ are chosen.}
\label{fig:example-conditional-dag-task-instance}
\end{figure}

For instance the distribution function of the conditional DAG illustrated 
in Figure~\ref{fig:example-conditional-dag-task} is devised by the calculation of the 
probability for each of the DAG's realizations and its respective parameter values.
The instance illustrated in Figure~\ref{fig:example-conditional-dag-task-instance} represents 
the graph where both upper edges are chosen for which the probability is $0.7 \cdot 0.6 = 0.42$. 
The associated length is $12$ and the associated volume is $13$. 
By similar reasoning, choosing the edges with probability $0.7 \cdot 0.4$, $0.3 \cdot 0.6$, and $0.3 \cdot 0.4$ 
yield $0.28$, $0.18$ or $0.12$ realization probability of the associated DAG structures.
Calculating the volume and length of each of these realizations yields the data listed in 
Table~\ref{table:example-probabilities}.
Consequently, we derive $F_i(u, v) = \mathbb{P}(vol(G_i) \leq u, \len(G_i) \leq v)$ as follows: 
\begin{dmath*}
\mathds{1}(u-13) \cdot \mathds{1}(v-12)  \cdot 0.42 + \mathds{1}(u-14) \cdot \mathds{1}(v-13)  \cdot 0.18 + \mathds{1}(u-10) \cdot \mathds{1}(v-9)  \cdot 0.28 + \mathds{1}(u-11) \cdot \mathds{1}(v-11)  \cdot 0.12
\end{dmath*}
where $\mathds{1}$ denotes the step function, i.e., 
$\mathds{1}(x)$ is $1$ if $x\geq 0$ and $0$ otherwise. 
We note that for probabilistic conditional DAG tasks 
as presented, the CDF is a step function with finitely many steps. 
Moreover, we assume that the probabilities of DAG instances are independent.

\subsection{Tardiness}

Every job that misses its deadline must be handled by the system, 
i.e., a mechanism must be devised that decides the actions taken upon such events. 
A common mechanism is the immediate abortion of every job which exceeds its deadline 
in order to avoid any interference of subsequent jobs. 
This approach is inefficient in the sense that all computation results 
and state changes are dumped and even may have to be revoked for 
consistency reasons, which holds especially true if the amount of 
time that the deadline is exceeded is rather small. 
Informally speaking, the tardiness of a job measures the delay of job 
with respect to its deadline.

\begin{definition}[Tardiness]
	\label{def:tardiness}
	Let $\delta_i(\ell)$ denote the tardiness of the $\ell$-th job of task 
	$\tau_i$, i.e., the amount of time that the $\ell$-th job exceeds the task's deadline 
	under the consideration of possibly pending workload from prior jobs.
        The tardiness can be recursively stated as $\delta_{i}(\ell) =  \max\{\delta_{i}(\ell-1)+(R_{i,\ell}-D_i),0\}$,
        where $R_{i,\ell}$ denotes the response time of the $\ell-th$ job of task $\tau_i$.
        Furthermore $\delta_{i}(0) = 0$ by definition.
	\hfill \IEEEQED
\end{definition}
We note that due to this definition, the $\ell$-th job of task $\tau_i$ does meet its deadline if $\delta_{i}(\ell) = 0$, 
and it does miss its deadline if $\delta_{i}(\ell) > 0$.
In pursuance of improving this problem we intent to bound 
the tardiness of each job of a task by a tardiness bound. 

\begin{definition}[Tardiness Bound]
	\label{def:tardiness-bound}
	A task $\tau_i$ is said to have a tardiness bound $\rho_i>0$ if any job 
	of that task will be aborted if the job's tardiness exceeds $\rho_i$, 
	i.e., we have $0 \leq \delta_i(\ell) \leq \rho_i$ for all $\ell \geq 0$.
	\hfill \IEEEQED
\end{definition}

The tardiness bound is user-specified and refines the formal description of 
a probabilistic sporadic constrained-deadline parallel DAG task to the tuple $(F_i, D_i, T_i, \rho_i)$.

\subsection{Deadline Misses}

We pursue to design reservation systems that provide sufficient service 
to each task $\tau_i$ in the task set \mbox{$\mathbb{T} = \{\tau_1, \tau_2, \ldots, \tau_n\}$} 
such that the probability of $k$ consecutive deadline misses is bounded.

\begin{definition}[Consecutive Deadline Misses]
\label{def:consecutive-deadline-misses}
 Any sequence of $k$ consecutive job releases $\tau_{i, \ell}, \tau_{i, \ell+1}, \ldots, \tau_{i, \ell+k-1}$ 
 for $\ell \geq 0$ is subject to $k$-consecutive deadline misses if 
 the following conditions hold:
 \begin{itemize} 
   \item All jobs in the sequence miss their deadline
   \item Either $\ell=0$ or the previous job $\tau_{i,\ell-1}$ does not miss its deadline.\hfill \IEEEQED
\end{itemize}
\end{definition}

For each task we define a function $\theta_i : \mathbb{N} \to [0,1]$ to specify 
that we tolerate $k$ consecutive deadline misses 
for a given probability of at most $\theta_i(k)$.

\begin{definition}[$k$ Consecutive Deadline Miss Constraint]
\label{def:k-consecutive-deadline-constraints}
 Let $\phi_i(j, k) := \mathbb{P}(\delta_{i}(j)>0, \dots, \delta_{i}(j+k-1)>0~|~j=0 \text{ or }\delta_{i}(j-1)=0 )$ denote the probability that 
 the sequence $\tau_{i, j}, \tau_{i, j+1}, \ldots, \tau_{i, j+k-1}$ suffers from 
 $k$-consecutive deadline misses.
 Then a probabilistic conditional 
 DAG task $\tau_i$ is said to satisfy the deadline constraint $\theta_i(k)$ 
 if
 \begin{equation}\label{eq:cons_deadl_miss_constr}
 	\sup_{j \geq 0} \{\phi_i(j, k)\} = \phi_i(0, k) \leq \theta_i(k),
 \end{equation}
i.e., at each position $j$ the probability $\phi_i(j,k)$ does not exceed the threshold $\theta_i(k)$.
\hfill \IEEEQED
\end{definition}
We note that the equality in Eq.~\eqref{eq:cons_deadl_miss_constr} is due to the lack of pending workload prior to the release of job $\tau_{i,j}$.

\section{Scheduling Problem}
\label{sec:scheduling-problem}

We use a reservation system to handle the scheduling of 
the DAG tasks and use any partitioned scheduling algorithm to schedule the 
reservation system and other tasks in the system.

\subsection{Reservations}
\label{sec:reservations}

In a \emph{reservation system} service is reserved for each probabilistic parallel DAG task $\tau_i$ due to some regulation. 
At those reservations the task instances of $\tau_i$ can be processed.
The reservation system is $m_i$-\emph{in-parallel} if there are at most $m_i\in \mathbb{N}$ reservations at the same time.
In this work we consider a simplified version of in-parallel reservation system:

\begin{definition}[Our Reservation System]
	\label{def:in-parallel_reservation}
	A \emph{reservation system} consists of $m_i$ reservation servers that provide 
	$E_i$ amount of service each and that is replenished every $P_i >0$ time units. 
	More specifically, to provide the service, each $P_i$ time units there are activated a 
	multiset of $m_i\in \mathbb{N}$ distinct reservations, that each guarantee a service of $E_i$ time 
	units over an interval of length $P_i$.
\end{definition} 

The instances of a task are assigned to the provided service in first-in-first-out (FIFO)-manner.
Furthermore, we assume that at each time all assigned reservations only serve the subjobs of a single DAG job by the FIFO-policy.
The reservation system is scheduled upon $M$ identical multiprocessors 
according to any scheduling paradigm and provides service to the DAG jobs 
whenever they are scheduled as follows.

\begin{definition}[List-Scheduling]
	\label{def:list-scheduling} 
	In a list schedule on $m_i$ in-parallel reservation servers  
	a subjob of a given DAG job $G = (V, E)$
	is executed on any reservation server that is idle and 
	scheduled for execution and as soon as all preceding subjobs 
	have executed until completion.
	More formally, the starting time $s_{i}$ for each subjob $v_{i}$ 
	is given by $\min\{t~|~\text{some scheduled reservation server idles at }t,~t\geq \max\{f_{j}~|~v_{j} \in pre(v_{i}) \}\}$.
	\hfill \IEEEQED
\end{definition}

For the remainder of this section, we assume the existence of a \emph{feasible} 
schedule $S$ upon $M$ identical multiprocessors, 
meaning that all reservations will 
provide the promised service.

\begin{definition}[Work]
	Let $work_i^S(t_1, t_2)$ denote the amount of workload from DAG jobs derived by task $\tau_i$ that was \emph{worked} during the time interval $t_1$ to $t_2$ given the schedule $S$.
\hfill \IEEEQED
\end{definition}

Based on this definition, the worst-case response time of a job $\tau_{i,\ell}$ of a DAG task $\tau_i$
that was released at $t_{i, \ell}$ is given by the smallest $t'\geq t_{i, \ell}$ such that 
$work_i^S(t_{i,\ell}, t') \geq vol(G_i^{\ell}) + backlog_i^S(t_{i, \ell})$, where $backlog_i^S(t_{i, \ell})$ is the amount of unfinished work at time $t_{i,\ell}$ of jobs of $\tau_i$ released before $t_{i,\ell}$. 
Note that $backlog_i^S(t_{i, \ell}) = 0$ if there are no previous
deadline misses since we assume $D_i\leq P_i$ in our system model.
In the following we express the processed work in terms of provided service and develop a response-time bound as stated in Theorem~\ref{thm:first_response-time_bound}.

For sake of argument, let $S$ denote a \emph{feasible} schedule of a reservation system 
that works a job of a DAG task $\tau_i$ until completion.
Furthermore let $serv_i^S(t_1, t_2)$ denote the service that is 
provided to the DAG job during the time interval from $t_1$ to $t_2$ 
in the schedule $S$.

\begin{definition}[Envelope]
	Let $S$ be a concrete schedule of $\mathbb{T}$.
	Consider a given DAG job instance $G$ of some task in $\mathbb{T}$ with subjobs $V=\setof{v_1, \dots, v_\ell}$.
	Let each subjob $v_k$ have the starting time $s_k$ and finishing time $f_k$ in $S$.
	We define the envelope $s_{k_1}, f_{k_1}, s_{k_2}, f_{k_2}, \dots,  s_{k_p}, f_{k_p}$ of $G$, with $p\in \{1, \dots, \ell\}$, recursively by the following properties:
 \begin{enumerate}
 	\item $k_{i} \neq k_j \in \setof{1,\dots,\ell}$ for all $i\neq j$
 	\item $v_{k_p}$ is the subjob of $V$ with maximal finishing time
 	\item $v_{k_{i-1}}$ is the subjob in $pre(v_{k_i})$ with maximal finishing time, for all $i \in \setof{p, p-1, \dots, 2}$
 	\item $pre(v_{k_1}) = \emptyset$ 
 \end{enumerate}
We note that the definition of an envelope for a DAG job instance may be not unique if there are subjobs with equal finishing time.
In this case we choose one among them arbitrarily.
\hfill \IEEEQED
\end{definition}

Based on the definition of an envelope, 
we are able to formally state the following lemma.

\begin{lemma}
\label{lemma:least-worked}
	Given a schedule $S$ of $\mathbb{T}$.
	We consider a task $\tau_i \in \Tbb$ with an $m_i$-in-parallel reservation system.
%	 reservation system as described in Definition~\ref{def:in-parallel_reservation}.
	Let $G=\tau_{i,j}$ be one DAG job instance of $\tau_i$ with envelope 
	$s_{k_1}, f_{k_1}, \dots,  s_{k_p}, f_{k_p}$.
	Then the amount of work that is finished during the interval from $f_{k_{q-1}}$ to $f_{k_{q}}$ for $q\in \{2, \dots, p\}$ is lower bounded by 
	\begin{align*}
		work_i^S(f_{k_{q-1}}, f_{k_{q}}) \geq & serv_i^S(f_{k_{q-1}}, s_{k_{q}}) + serv_i^S(s_{k_{q}}, f_{k_{q}}) \\&- (m_i-1) \len(v_{k_{q}})
	\end{align*}
	where $v_{k_{q}}$ is the subjob from the envelope starting at time $s_{k_{q}}$ and finishing at $f_{k_{q}}$.
\end{lemma}

\begin{IEEEproof}
	In the proof we split the work at time $s_{k_q}$ and estimate each summand of $work_i^S(f_{k_{q-1}}, f_{k_q}) = work_i^S(f_{k_{q-1}}, s_{k_q}) + work_i^S(s_{k_q}, f_{k_q})$ on its own.
	Combining both estimations yields the desired result.
	
 In a first step we will prove that between finish and start of two consecutive subjobs in the envelope, the provided service is fully utilized by the DAG instance, i.e.,
 \[work_i^S(f_{k_{q-1}}, s_{k_q}) = serv_i^S(f_{k_{q-1}}, s_{k_q})\] holds for all $q \in \{2, \dots, p\}$. 
 Given the workload conserving properties of list-scheduling used 
 to dispatch subjobs to the service, an eligible subjob is scheduled 
 whenever service is available. 
 Since by definition $s_{k_q}$ is the earliest time that $v_{k_q}$ 
 is able to execute, all service during $f_{k_{q-1}}$ to $s_{k_q}$ must 
 have been used to \emph{work} on other (non envelope) subjobs.
 
 Secondly, we show that the workload $work_i^S(s_{k_q}, f_{k_q})$ from start to finish of a subjob in the envelope can be estimated by
 \[\max \{ serv_i^S(s_{k_q}, f_{k_q}) - (m_i-1)  \cdot \ell en(v_{k_q}) , \ell en(v_{k_q}) \}.\]
 Clearly, during the starting time and finishing time of $v_{k_q}$ at least $\ell en(v_{k_q})$ will be worked.
 Additionally, given the provided service $serv_i^S(s_{k_q}, f_{k_q})$ 
 due to sequential execution of $v_{k_q}$, at most $m_i-1$ reservations of 
 duration $\ell en(v_{k_q})$ may be unused. 
 Therefore $work_i^S(s_{k_q}, f_{k_q}) \geq \max \{ serv_i^S(s_{k_q}, f_{k_q}) - (m_i-1)  \cdot \ell en(v_{k_q}) , \ell en(v_{k_q}) \}$. 
%\mario{If we do not use this max, we should delete it from the proof}
\end{IEEEproof}

Based on this lemma, we can calculate the response-time of a DAG job.
To do this we first extend the Lemma.

\begin{lemma}
	\label{lem:least-worked-extended}
	Under the conditions of Lemma~\ref{lemma:least-worked}, we have that
	\begin{equation}
		work_i^S(r_G, r_G+t) \geq serv_i^S(r_G, r_G+t) - (m_i-1)\len(G)
	\end{equation} holds, where $r_G$ is the release of job $G$ and $0\leq t \leq f_{k_p}$.
\end{lemma}

\begin{IEEEproof}
	The main part to prove this lemma is already done in Lemma~\ref{lemma:least-worked}.
	We just have to be careful about the scenarios where $t$ is not a time instant of the envelope.
	
	Similarly to the proof of Lemma~\ref{lemma:least-worked} we can show that $work_i^S (f_{k_{q-1}}, t) = serv_i^S (f_{k_{q-1}}, t)$ for all $t\in [f_{k_{q-1}}, s_{k_{q}}]$ and that $work_i^S ( s_{k_q}, t) \geq serv_i^S ( s_{k_q}, t) - (m_i-1) \len (v_{k_q})$ for all $t \in [s_{k_{q}}, f_{k_{q}}]$.
	Furthermore, by the same reasoning $work_i^S (r_G, t) = serv_i^S (r_G, t)$ holds for all $t\in [r_G, s_{k_{1}}]$.
	
	We obtain the desired result by splitting the interval $[r_G, t]$ into parts already described above and estimating all of them at the same time. 
	To formalize this, we define \[\mu \eqdef (r_G, s_{k_1}, f_{k_1}, \dots, s_{k_p}, f_{k_p}).\]
	For $q\in \{1, \dots, 2p+1\}$ we denote by $\mu(q)$ the $q$-th entry of $\mu$ and by $\mu^t(q) \eqdef \min\{\mu(q),t\}$ the $q$-th entry bounded by $t$.
%	{\footnotesize{}
%	\begin{align*}
%		&work_i^S(r_G, r_G+t)
%		\\&=\sum_{q=1}^{2p} work_i^S (\mu^t(q), \mu^t(q+1) )
%		\\& = \sum_{q=1}^{p} work_i^S (\mu^t(2q-1), \mu^t(2q) ) + \sum_{q=1}^{p} work_i^S (\mu^t(2q), \mu^t(2q+1) )
%		\\&\geq \sum_{q=1}^{p} serv_i^S (\mu^t(2q-1), \mu^t(2q) ) \\&\phantom{\geq{}}+ \sum_{q=1}^{p} \left(serv_i^S (\mu^t(2q), \mu^t(2q+1) ) - (m-1) \len(v_{k_q}) \right)
%		\\&= \sum_{q=1}^{2p} serv_i^S (\mu^t(q), \mu^t(q+1) ) - (m-1)\left( \sum_{q=1}^{p} \len(v_{k_q}) \right)
%		\\& \geq serv_i^S(r_G,r_G+t) - (m-1) \len(G)
%	\end{align*}
%	}
%
%	\mario{

	By decomposing $work_i^S(r_G, r_G+t)$, we obtain that it can be written as the sum of $\sum_{q=1}^{p} work_i^S (\mu^t(2q-1), \mu^t(2q) )$ and of $\sum_{q=1}^{p} work_i^S (\mu^t(2q), \mu^t(2q+1) )$.
	The first summand is lower bounded by the sum of the corresponding service values $\sum_{q=1}^{p} serv_i^S (\mu^t(2q-1), \mu^t(2q) )$, and the second summand from above is lower bounded by $\sum_{q=1}^{p} \left(serv_i^S (\mu^t(2q), \mu^t(2q+1) ) - (m-1) \len(v_{k_q}) \right)$.
	By combining both of the results, we obtain the lower bound
	\[\sum_{q=1}^{2p} serv_i^S (\mu^t(q), \mu^t(q+1) ) - (m-1)\bigg( \sum_{q=1}^{p} \len(v_{k_q}) \bigg),\] which is again bounded by $serv_i^S(r_G,r_G+t) - (m-1) \len(G)$.
	We conclude that $work_i^S(r_G, r_G+t)\geq serv_i^S(r_G,r_G+t) - (m-1) \len(G)$.
%	}
\end{IEEEproof}

\begin{definition}[Service Bound Function]
	For a task $\tau_i \in \Tbb$ the minimal service that is provided by the reservation system during an interval of length $t\geq 0$ is denoted by $sbf_i(t)$.
	We call $sbf_i$ the \emph{service bound function} of $\tau_i$.
\hfill \IEEEQED
\end{definition}

We use the service bound function to provide a lower bound $serv_i^S(r_G, r_G+ t) \geq sbf_i(t)$ for all schedules $S$. 
This leads us to the following theorem.

\begin{theorem}[Response-Time Bound]
	\label{thm:first_response-time_bound}
	We consider a task $\tau_i \in \Tbb$.
	Assume that the reservation system of $\tau_i$ is $m_i$-in-parallel and its minimal service is described by $sbf_i$.
	Let $G$ be the DAG which describes the task instance $\tau_{i,j}$ of $\tau_i$.
	Then the response time of $G$ is upper-bounded by
	\begin{equation}
		\min\{t > 0 ~|~ sbf_i(t) \geq W_i^G\}. 
	\end{equation}
	where $W_i^G \eqdef vol(G) + (m_i-1) \cdot \ell en(G) + backlog_i^S(r_G)$ for notational brevity
\end{theorem}

\begin{IEEEproof}
	Let $t' \eqdef \min\{t > 0 ~|~ sbf_i(t) \geq W_i^G\}$.
	We do the proof by contraposition:
	If we assume that $t'$ does not bound the response time, then $t'< f_{k_p}$, where $f_{k_p}$ is the last entry in the envelope of $G$.
	In this case Lemma~\ref{lem:least-worked-extended} yields:
	\begin{align*}
		work_i^S(r_G, r_G+t') 
		&\geq serv_i^S(r_G, r_G+t') - (m_i-1)\len(G)
		\\&\geq  sbf_i(t') - (m_i-1)\len(G)
	\end{align*}
	By the definition of $t'$ we have $sbf_i(t') \geq vol(G) + (m_i-1) \cdot \ell en(G) + backlog_i^S(r_G)$.
	Hence, 
	\begin{equation*}
		work_i^S(r_G, r_G + t') \geq vol(G) + backlog_i^S (r_G)
	\end{equation*}
	the job $G$ is finished at time $t'$, i.e., $t'\geq f_{k_p}$.
\end{IEEEproof}

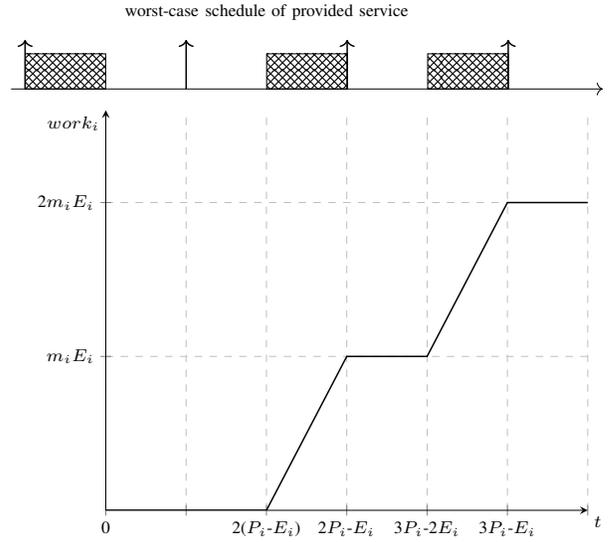
\begin{figure}[tb]
\centering
\resizebox{.45\textwidth}{!}{\pgfplotsset{%
    ,compat=1.12
    ,every axis x label/.style={at={(current axis.right of origin)},anchor=north west}
    ,every axis y label/.style={at={(current axis.above origin)},anchor=north east}
    }

\begin{tikzpicture}
\scriptsize{}

%VARIABLES
\pgfmathsetmacro{\x}{6.87/6}
\pgfmathsetmacro{\dx}{3.1}

% Now use \PHI anywhere you want -15 to appear,
% can also be used in calculations like 2*\PHI

\node at (2*\x,7.1) {worst-case schedule of provided service};

\draw[->] (-\x-0.2,6) -- (\x*6+0.2,6);
\draw[->, semithick] (-\x,6) -- (-\x,6.7);
\draw[->, semithick] (\x,6) -- (\x, 6.7);
\draw[->, semithick] (3*\x,6) -- (3*\x,6.7);
\draw[->, semithick] (5*\x,6) -- (5*\x,6.7);

\draw[pattern=crosshatch] (-\x,6) rectangle (0,6.5);
\draw[pattern=crosshatch] (2*\x,6) rectangle (3*\x,6.5);
\draw[pattern=crosshatch] (4*\x,6) rectangle (5*\x,6.5);

\begin{axis}[%
    ,xlabel=$t$
    ,ylabel=$work_i$
    ,axis x line = bottom,axis y line = left
    ,ytick={0.25, 0.5}
    ,xtick={0, 2, 4, 6, 8, 10, 12}
    ,xticklabel style={
                        %rotate=20
}
    ,ymax=0.65 % or enlarge y limits=upper
    ,yticklabels={$m_i E_{i}$, 2$ m_i E_{i}$} 
    ,xticklabels={0,,2($P_i$-$E_i$), 2$P_i$-$E_i$ ,3$P_i$-2$E_i$, 3$P_i$-$E_i$} 
    ,grid=both,
    ,grid style={line width=.1pt, dashed}
    ]

\addplot+[sharp plot, no marks, semithick, black] coordinates {(0,0) (4,0) (6,0.25) (8, 0.25) (10, 0.5) (12,0.5)};
%\addplot+[sharp plot, no marks, thin, dashed, black] coordinates {(0,0) (2.0*\dx,0) (4.0*\dx, 0.25) (6.0*\dx, 0.5)} node[below right, pos=.5,black] {$sbf_i(t)'$};
\end{axis}
\end{tikzpicture}}
\caption{Supply Bound Function $sbf(t)$ of the reservation system.}
\label{fig:supply-bound-function}
\end{figure}

We emphasize that the reservation schemes and respective supply-bound function 
are not enforced to follow any specific kind of reservation scheme. 
The complexity of the calculation of the response-time depends only on the 
supply bound function. 
For instance, Figure~\ref{fig:supply-bound-function} shows the supply-bound function 
of a \emph{our reservation system} from Definition~\ref{def:in-parallel_reservation}. As depicted, there may be no service provided to the task 
for up to $2(P_i-E_i)$ time units in the worst case.
%, e.g., when the service occurs at 
%the last possible time.
%
We note that the first activation of reservations has to occur no later than at the release of the first job of $\tau_i$. 
Otherwise our analysis becomes invalid.
However, the reservation system can stop assigning new reservation servers if there is no pending or 
unfinished job of $\tau_i$, as long as it starts assigning new reservations if new jobs arise in the ready queue.

If we assume a reservation server as in Definition~\ref{def:in-parallel_reservation}, then the response-time or service-time of a DAG job $G$ is described by the 
following theorem.

\begin{theorem}[Service Time]
\label{thm:service-time}
	Let $G = \tau_{i,j}$ be a task instance of $\tau_i$.
	We assume that for $\tau_i$ we have a reservation system as in Definition~\ref{def:in-parallel_reservation} 
        with $m_i$ equal sized in-parallel services $E_i\leq P_i$.
	We can give an upper bound $R_G$ on the response time of $G$ by
	\begin{equation}\label{eq:R_G}
		R_G = \left(\ceiling{\frac{W_i^G}{m_i E_i}} +1 \right) (P_i-E_i) + \frac{W_i^G}{m_i}
	\end{equation}
	where $W_i^G \eqdef vol(G) + (m_i-1)\len(G)+ backlog_i^S(r_G)$ for notational brevity.
\end{theorem}

\begin{IEEEproof}
For the proof we assume that $vol(G)>0$ since otherwise no work has to be done and $R_G = 0$ is already a trivial response-time bound.
We aim to utilize Theorem~\ref{thm:first_response-time_bound}.
Therefore, we have to find the minimal $t > 0$ such that $sbf_i(t) = W_i^G$.
In the following we show one illustrative and one formal proof to justify that this minimal $t$ is in fact $R_G$ from Eq.~\eqref{eq:R_G}:

We assume the worst-case service as depicted in Figure~\ref{fig:supply-bound-function}.
We can see in the figure that every time when service is provided, it is done on $m_i$ resources simultaneously.
Hence, the total time which $\tau_i$ has to be served, until $G$ is finished, is $\frac{W_i^G}{m_i}$.
This happens during $\ceiling{\frac{W_i}{m_i \cdot E_i}} + 1$ service cycles. 
Therefore, we have to add this many times the amount of the service cycle, where $\tau_i$ is not served, i.e., $(P_i-E_i)$.
In total, the response time is 
\begin{math}
\left( \ceiling{\frac{W^G_i}{m_i\cdot E_i}} + 1 \right) (P_i-E_i)
+ \frac{W^G_i}{m_i}.
\end{math}

For the more formal proof, we also assume the worst-case service from Figure~\ref{fig:supply-bound-function}.
For the function $g: \mathbb{R}_{>0} \to \mathbb{R}_{>0}$ with 
\[g(t) \eqdef \left(\ceiling{\frac{t}{m_i E_i}} +1 \right) (P_i-E_i) + \frac{t}{m_i} \]
the composition $sbf\circ g$ is the identity and the function $g$ picks the minimal value of the inverse image of $sbf_i(t)$, i.e., \mbox{$g(t) = \min(sbf_i^{-1}(t))$} holds.
Hence, we obtain \mbox{$g(W_i^G) = \min \{t>0 ~|~ sbf_i(t) \geq W_i^G\}$}.
\end{IEEEproof}

In general, if we know an upper bound $b$ on the backlog of the previous job, we can state the response time bound from Eq.~\eqref{eq:R_G} independent from the previous schedule, by 
\begin{equation}\label{eq:R_G'}
	R'_G(b) = \left(\ceiling{\frac{V_i^G(b)}{m_i E_i}} +1 \right) (P_i-E_i) + \frac{V_i^G(b)}{m_i}
\end{equation}
where $V_i^G(b) \eqdef vol(G) + (m_i-1)\len(G)+ b$.
Based on Eq.~\eqref{eq:R_G'}, we
bound the response time for the case that the preceding job has a deadline miss and for the case that the preceding job has \emph{no} deadline miss.
\begin{cor}\label{cor:response_time_bound}
	Under the assumptions of Theorem~\ref{thm:service-time},
	\mbox{$R'_G(\rho_i \cdot m_i)$} is an upper bound on the response time of $G$ if the preceding job has a deadline miss, and $R'_G(0)$ is an upper bound if the preceding job has no deadline miss.
\end{cor}

\begin{IEEEproof}
	This follows directly from Theorem~\ref{thm:service-time} by using either $backlog_i^S(r_G) \leq \rho_i \cdot m_i$ (in case of a deadline miss) or $backlog_i^S(r_G) = 0$ (in case of \emph{no} deadline miss).
\end{IEEEproof}

\section{Reservation Analysis and Optimization}
\label{sec:probabilistic-reservation-analysis-and-design}

In this section we devise the analysis and optimization algorithm to 
generate reservation systems that provably respect the upper-bounds for $k$ consecutive deadline misses 
in a probabilistic sense. We emphasize that in order to co-design the $k$ consecutive deadline-miss 
constraints with the reservations configurations time-efficient algorithms are required 
to calculate the probabilities for $k$ consecutive deadline misses for any given 
reservation configuration.

\subsection{Analysis of Reservation Systems}
Based on the finite sample space of DAG structures $G$ of the probabilistic 
conditional DAG tasks $\tau_i$ we define the random variables $R_i^1 \eqdef (G \mapsto R'_G(\rho_i m_i))$ and $R_i^0 \eqdef (G \mapsto R'_G(0))$, which yield
for each DAG job the response time bounds from Corollary~\ref{cor:response_time_bound} with and without a previous deadline miss. 
According to Definition~\ref{def:k-consecutive-deadline-constraints}, the constraint for $k$ consecutive deadline misses is fulfilled if
\begin{equation}
	\phi_i(0, k) \leq \theta_i(k),
\end{equation}
where $\phi_i(0, k)$ is the probability that the first $k$ jobs of $\tau_i$ miss their deadline, and $\theta_i(k)$ is some predefined value.

Since $\phi(0,k) = \mathbb{P}\left(\delta_{i}(k) > 0, \delta_{i}(k-1) > 0, \ldots, \delta_{i}(1) > 0 \right)$, we can use Bayes' Theorem, to reformulate $\phi(0,k)$ as
\begin{equation*}
	\mathbb{P}\left(\delta_{i}(k) > 0~|~\delta_{i}(k-1) > 0, \ldots, \delta_{i}(1) > 0\right) \cdot \phi_i(k-1).
\end{equation*}
The probability that $\tau_{i,k}$ does not meet its deadline does not decrease if the tardiness of the preceding job is increased.
Therefore, if $\delta_i(k-1) = \rho_i$, then the probability for a deadline miss of $\tau_{i,k}$ is maximal.
In this case, the amount of tardiness of the other jobs $\delta_i(k-2), \dots, \delta_i(1)$ is irrelevant for the tardiness of $\tau_{i,k}$.
More specifically, 
	\begin{equation*}
		\begin{split}
		&\mathbb{P}\left(\delta_{i}(k) > 0~|~\delta_{i}(k-1) > 0, \ldots, \delta_{i}(1) > 0\right) 
		\\&\qquad \leq \mathbb{P} \left( \delta_i(k)>0~|~\delta_{i}(k-1) = \rho_i \right)
		\end{split}
	\end{equation*}
	holds and we can thus bound the probability for $k$ consecutive deadline misses by 
	\begin{equation}
		\phi_i(0,k) \leq \mathbb{P} \left( \delta_i(k)>0~|~\delta_{i}(k-1) = \rho_i \right) \cdot \phi_i(0,k-1).
	\end{equation}
	Then by Corollary~\ref{cor:response_time_bound} we know that 
	\begin{align*}
		\mathbb{P} \left( \delta_i(k)>0~|~\delta_{i}(k-1) = \rho_i \right)  \leq \mathbb{P} \left( R^1_i > D_i \right)
	\end{align*}
	and for the probability of the first job (without previous deadline miss)
	\begin{align*}
		\phi_i(0, 1) = \mathbb{P} \left( \delta_i(1)>0 \right) \leq \mathbb{P} \left( R^0_i > D_i \right).
	\end{align*}
	Combining the results yields a bound on the probability of $k$ consecutive deadline misses:
	\begin{align*}
		\phi_i(0, k) 
		& \leq \mathbb{P} \left( R^1_i > D_i \right) \cdot \phi_i(0, k-1)
%		\\&\leq \mathbb{P} \left( R'_i > D_i \right)^2 \cdot \theta_i(k-2)
		\\&\leq \dots
		\leq \mathbb{P} \left( R^1_i > D_i \right)^{k-1} \cdot \phi_i(0, 1)
		\\&\leq \mathbb{P} \left( R'_i > D_i \right)^{k-1} \cdot \mathbb{P} \left( R^0_i > D_i \right)
%		\\&\leq \mathbb{P} \left( R'_i > D_i \right)^{k}
	\end{align*}
Since $\mathbb{P} \left( R^0_i > D_i \right) \leq \mathbb{P} \left( R^1_i > D_i \right)$, we also derive a simplified bound for the probability of $k$ consecutive deadline misses 
of task $\tau_i$ by 
\begin{equation}
 \label{eq:deadline-bound}
	\phi_i(0, k)
	\leq \mathbb{P} \left( R^1_i > D_i \right)^{k}.
\end{equation}
As a prerequisite to derive upper-bounds on response-times for queuing systems 
it must be shown that the system is stable. Informally speaking this means that 
all backlog of the reservation system will have been worked at some point in time. 
We first give a formal definition of stability and then show that our devised 
reservation-based queuing system is stable by construction.

\begin{definition}[Stability]
	\label{def:stability}
	A reservation system $\mathcal{R}_i$ is considered \emph{stable} if 
	for all $\ell \geq 0$ with $\delta_i(\ell) = 0$ it is almost certain that 
	there exists $k > 0$ such that $\delta_i(k+\ell) = 0$.
	More formally,
	\begin{equation}
		\lim_{k\to\infty} \phi(0,k) =0,
	\end{equation}
	i.e., the probability for $k$ consecutive deadline misses approaches $0$ for $k\to\infty$.
	\hfill \IEEEQED
\end{definition}

\begin{theorem}[Stability]
\label{thm:stability-theorem}
 A reservation system $\mathcal{R}_i$ is stable if 
 \mbox{$\mathbb{P}(R'_i>D_i)<1$}. 
\end{theorem}

\begin{IEEEproof}
	The probability for $k$ consecutive deadline misses is bounded by $\phi_i(0,k) \leq \mathbb{P} \left( R^1_i > D_i \right)^{k}$ according to Eq.~\eqref{eq:deadline-bound}.
	If $\left( R^1_i > D_i \right)<1$, then $\mathbb{P} \left( R^1_i > D_i \right)^{k} \to 0$ for $k \to\infty$.
	This concludes the theorem.
\end{IEEEproof}

In consequence we do not have to especially consider stability 
concerns in the design of the reservation systems other than 
$k$-consecutive deadline constraints.

\subsection{Distribution Function Calculation }

In this section, we show how to practically calculate 
the response-time upper bounds.
First, we define the auxiliary random variable 
\[
X_i \eqdef \frac{vol(G) + (m_i-1) \cdot \len(G)+ \rho_i \cdot m_i}{m_i \cdot E_i} = \frac{V_i^G}{m_i E_i}\]
for which the distribution function $\mathbb{P}(X_i \leq u)$ can be directly 
computed from the probabilistic DAG task model, i.e., by enumerating over all 
possible DAG job structures weighted by their realization probabilities 
as previously described. 
With reference to Corollary~\ref{cor:response_time_bound}, 
the distribution function of $R^1_i$ can be written as follows:
\[
 \mathbb{P}(R^1_i \leq u)   = \mathbb{P}\left((P_i-E_i) \cdot (\ceiling{X_i}+1) + E_i \cdot X_i \leq u\right)
\]
Let $dom(X_i)$ denote all values that $X_i$ can take, then we define the set of 
constant values $I_i \eqdef \{ \ell \in \mathbb{N}~|~\floor{\inf(dom(X_i))} \leq \ell \leq \ceiling{\sup(dom(X_i))} \}$.
Moreover given $I_i$ the domain of $\psi(X_i) = (P_i-E_i) \cdot (\ceiling{X_i}+1) + E_i \cdot X_i$ 
can be partitioned as follows:
\[ 
 \bigcup_{\ell \in I_i} \{(P_i-E_i) \cdot (\ell+2) + E_i \cdot X_i~|~\ell < X_i \leq \ell + 1 \}
\]
by the fact that $\ceiling{X_i} \mapsto \ell+1$ for every $X_i \in (\ell, \ell+1]$. 
By the $\sigma$-additivity property of distribution functions and rearrangements yields
\begin{equation}
\label{eq:response-time-distribution}
 \sum_{\ell \in I_i} \mathbb{P}(X_i \leq \frac{u-(P_i-E_i)\cdot (\ell+2)}{E_i}~|~\ell < X_i \leq \ell+1)
\end{equation}

\subsection{Optimization of Reservation Systems}
\label{sec:calculation-of-reservation-systems}

\begin{algorithm}[tb]
\caption{Calculation of Reservation Systems}
\begin{algorithmic}[1]
\Require{$\mathbb{T},~\theta_1(k_1), \theta_2(k_2), \ldots, \theta_n(k_n),~\Omega_1, \Omega_2,  \ldots, \Omega_n$;}
\Ensure{$\mathcal{R}_1, \mathcal{R}_2, \ldots, \mathcal{R}_n$ that satisfy the above requirements;}
\State{Initialize reservations $\mathcal{R} \leftarrow \{\}$;}
\For{each task $\tau_i$ in $\{ \tau_1, \tau_2, \ldots, \tau_{n}\}$} 
 \For{$m_i$ in $\{1, 2, \ldots, \Omega_i \}$}
   \State{$E_i \leftarrow \min \{E_i~|~(\Phi^n_i)^{k_i} \leq \theta_i(k_i) \}$;}
   \If{$E_i$ could not be found}
     \State continue;
   \Else
     \State{$\mathcal{R}_i \leftarrow \mathcal{R}_i \cup \{ m_i$ reservations with service $E_i\}$;}
   \EndIf
 \EndFor
\EndFor
\Return $\mathcal{R}$;
\end{algorithmic}
\label{alg:reservation-design-greedy}
\end{algorithm}

In this section we present Algorithm~\ref{alg:reservation-design-greedy} to calculate 
reservation systems for the scheduling of probabilistic constrained-deadline conditional
DAG tasks. Under the consideration of probabilities of upper-bounds for the maximal 
number of tolerable $k_i$ consecutive deadline misses and given tardiness bounds 
the objective is to find minimal numbers of in-parallel reservations $m_i$ and 
associated minimal amounts of service time $E_i$. 
For each probabilistic constrained-deadline conditional DAG task the algorithm 
determines all feasible configurations $(m_i, E_i)$ by iterating through the number of 
in-parallel reservations $m_i \in [1, \Omega_i]$ and search for the smallest
required reservation service to still comply with the consecutive deadline-miss 
constraints.

\begin{theorem}[Monotonicity]
\label{thm:service-time-monotonic}
The functions \[ \Phi^n_i: \mathbb{R}_{>0} \rightarrow \mathbb{R}_{>0},~E_i \mapsto \mathbb{P}(R^1_i > D_i)_{|m_i=n} \]
that yield the probabilities of an upper-bound of a deadline-miss for a fixed number of in-parallel 
reservations with respect to the service time $E_i$ are monotonically decreasing.
\end{theorem}

\begin{IEEEproof}
For easier readability let 
\[
 Y_i \eqdef \frac{vol(G_i)+(m_i-1) \cdot \ell en(G_i) + \rho_i \cdot m_i}{m_i}\]
for which the distribution function is independent of $E_i$ for every fixed $m_i$.
According to the definition of $\mathbb{P}(R^1_i > D_i)$ in the beginning of this section, 
we have to prove that 
\begin{align*}
& \mathbb{P}\bigg(\Big(\ceiling{\frac{Y_i}{E_i}}+1\Big) \cdot (P_i-E_i) + Y_i >D_i\bigg)   \\ &
\geq \mathbb{P}\bigg(\Big(\ceiling{\frac{Y_i}{E_i+\delta}}+1\Big) (P_i-(E_i+\delta)) + Y_i >D_i \bigg) 
\end{align*}
for any positive arbitrary increment $\delta \geq 0$ and any realizations of $Y_i \geq 0$.
Let an arbitrary realization $Y_i \geq 0$ satisfy
\[
(\ceiling{\frac{Y_i}{E_i+\delta}}+1) \cdot (P_i-(E_i+\delta)) + Y_i >D_i \]
In this case $Y_i$ satisfies
\[ (\ceiling{\frac{Y_i}{E_i}}+1) \cdot (P_i-E_i) + Y_i >D_i\] as well 
which yields the assumption by the property of distribution functions.
\end{IEEEproof}

Due to  the monotonicity of the functions $\Phi^n_i$ as shown in 
Lemma~\ref{thm:service-time-monotonic}, it is possible to find the minimal amount of 
reservation service to guarantee compliance with the consecutive deadline-miss 
constraints by using binary search in the interval $(0, D_i]$.
We emphasize that $\Omega_i$ is an upper-bound specified by the user that can be 
set to an arbitrary fixed number that is larger than the number of available processors or 
determined as the point where an increase in the number of 
in-parallel reservations does not yield a \emph{significant} decrease 
in the amount of required service to satisfy the deadline-miss probability 
constraints.
 
\section{Conclusion and Future Work}
\label{sec:conclusion-and-future-work}

In this paper we proposed a probabilistic version and formal description
of the widely used conditional parallel DAG task model and proposed a resource reservation 
system that allows for \emph{scheduling anomaly free} scheduling whilst provably 
guaranteeing probabilistic quantities such as bounded tardiness, stability, and probabilistic upper-bounds of 
$k$ consecutive deadline misses. 
In addition, we provided an algorithm to optimize the reservations systems 
with respect to the above quantities and showed that probabilistic conditional DAG tasks with a high degree 
of parallelism can improve a system's resource usage if deadline misses are allowed.
In the future we intent to improve the tightness of our proposed bounds and 
evaluate the effectiveness of the approach by implementing a prototype system.

\section*{Acknowledgments}
This work has been supported by European Research Council (ERC)
Consolidator Award 2019, as part of PropRT (Number 865170), and by
Deutsche Forschungsgemeinschaft (DFG), as part of Sus-Aware (Project
no. 398602212).

\bibliography{real-time}
\bibliographystyle{abbrv}

\end{document}